\documentclass[preprint]{aastex} 

\usepackage{graphicx}        
\usepackage{amssymb}         
\usepackage[usenames,dvipsnames]{color}          
\usepackage{epstopdf}
\definecolor{darkblue}{rgb}{0,0,0.5}
\usepackage{bm}
\DeclareFontFamily{OT1}{pzc}{}
\DeclareFontShape{OT1}{pzc}{m}{it}%
             {<-> s * [1.1500] pzcmi7t}{}
\DeclareMathAlphabet{\mathscr}{OT1}{pzc}%
                                 {m}{it}

\DeclareGraphicsExtensions{.pdf,.png,.jpg,.tif,.pdf}

\usepackage{amsmath}

\renewcommand{\a}{{\mathbf{a}}}

\newcommand{\g}{\bm{g}}
\newcommand{\z}{\mathbf{z}}

\newcommand{\x}{\mathbf{x}}
\newcommand{\A}{{\mathbf{A}}}
\newcommand{\B}{{\mathbf{B}}}

\newcommand{\T}{{\mathbf{T}}}

\newcommand{\grad}{\mbox{\boldmath$\nabla$}}

  \renewcommand{\le}{\leqslant}
  \renewcommand{\ge}{\geqslant}

\allowdisplaybreaks[1]

\begin{document}


\title{AN ANALYTICAL APPROACH TO SCATTERING BETWEEN TWO THIN MAGNETIC FLUX TUBES IN A STRATIFIED ATMOSPHERE}

\author{Chris S.~Hanson}
\author{Paul S.~Cally}

\affil{Monash Centre for Astrophysics and School of Mathematical Sciences,\\ Monash University, Clayton, Victoria 3800, Australia}
  \email{christopher.hanson@monash.edu}

\shortauthors{C.S. Hanson \& P.S. Cally}

\shorttitle{Scattering Between Two Stratified Tubes}

\begin{abstract}
We expand on recent studies to analytically model the behavior of two thin flux tubes interacting through the near- and acoustic far-field. The multiple scattering that occurs between the pair alters the absorption and phase of the outgoing wave, when compared to non-interacting tubes. We have included both the sausage and kink scatter produced by the pair. It is shown that the sausage mode's contribution to the scattered wave field is significant, and plays an equally important role in the multiple scattering regime. A disparity between recent numerical results and analytical studies, in particular the lack of symmetry between the two kink modes, is addressed. This symmetry break is found to be caused by an incorrect solution for the near-field modes.
\end{abstract}



\section{Introduction}
Magnetic features ranging from ensembles of slender tubes to large monolithic or spaghetti-like structures, are ubiquitous on the solar surface. Thin magnetic flux tubes, like those found in plage, interact strongly with the Sun's acoustic $p$-modes, both absorbing and scattering incident waves \citep{braun_etal_1990}. Whilst the mechanism for the observed absorption and scattering by large monolithic sunspots is fairly well understood, the similar (albeit smaller) observed affects of magnetic plage are not. Mechanisms operating within large structures are unlikely to be responsible for the absorption in smaller fibril structures like plage. Lately there has been a growing interest in developing a sound theoretical framework to model the collective behavior of adjacent flux tubes.

The absorption and scattering observed in plage is thought to be the result of a scattering regime between thin flux tubes present within the ensemble. Upon encountering a magnetic flux tube, a $p$-mode will lose some of its energy, which is deposited upon the tube, with the remaining energy scattered into the surrounding medium. The absorbed energy is converted into a vertically propagating slow wave \citep{cally_bogdan_1993,bogdan_etal_1996} manifesting itself as a kink or sausage like motion of the thin tube. In general, the scattered energy propagates away with a different phase. Within an ensemble, the scattered wave field will interact with nearby tubes and scatter again. This cascade of energy, as the waves scatter between the tubes is the multiple scattering regime \citep{bogdan_fox_1991}.
Initial attempts to model the multiple scattering regime were restricted to non-stratified media, due to the mathematical complexity that gravity introduces \citep[etc]{bogdan_zweibel_1987,keppens_etal_1994}. \citet{bogdan_zweibel_1987} were the first to attempt to characterize the multiple scattering regime, concluding that an incident wave's energy cascaded to smaller scales the further the wave travels into a fibril medium. \cite{bogdan_fox_1991} further shed light on the regime, finding that the scattered wave field from a pair of flux tubes will differ greatly from that of a single tube when the pair are within close proximity. Consequentially, the nature (monolithic or spaghetti) of the scatterer could then be discerned from the scattered wave field \citep{keppens_etal_1994}. However, by neglecting gravity, not only were the models simplified, they also neglected the powerful near-field resulting from stratification.

In the absence of gravity, the far field eigenfunctions suffice when matching internal and external wave fields. However, in a stratified atmosphere the downward propagating slow wave is also known to be an acoustic source, manifesting as a near-field acoustic jacket around the tube \citep{bogdan_cally_1995}. When considering compact acoustic sources, these jacket modes heal the surface velocity signature by removing the singularity present at $r=0$ (for axisymmetric cases) \citep{cally_2013}. These jacket modes, which form an infinite continuum, are radially evanescent but propagate up and down the tube. Consequentially, any neighboring tube will experience both the far- and near-field generated by a flux tube, and in turn respond to it, contributing its own scatter to the external medium. The scatter contributions  of all other tubes are no longer restricted to a discrete basis of far field modes, but rather consist of a combination of discrete far-field modes and a continuum of near field modes. The description of the acoustic jacket and its continuous spectrum in analytical models is challenging, and few scattering studies have included it. In the case of an isolated thin tube, the mathematical formalism has previously been developed for the dipole ($m=\pm1$), or kink \citep{hanasoge_etal_2008}, and monopole ($m=0$), or sausage \citep{hindman_jain_2012,andries_cally_2011}. Higher order fluting modes ($m\ge 2$) are not consistent with the thin flux tube approximation and will be ignored.

Scatter from an isolated tube is solely the product of the incident wave, and will only scatter into the incident wave's azimuthal order ($m$). However, in the presence of nearby tubes, the scatter from all tubes must be calculated simultaneously as each tube will contribute to its neighbor's scatter. To add to this challenge, the scatter is no longer bound to the $m$ mode of the incident wave and can now scatter into other $m$ modes. Basing their theory on \citet{kagemoto_yue_1986}, \citet{hanasoge_2009} were the first to analytically describe the interaction between a pair of thin flux tubes in a stratified atmosphere. They focused on the interaction of the kink mode oscillations $m=\pm 1$, ignoring the sausage mode ($m=0$) due to unsatisfactory boundary conditions, and concluded that the scatter properties of the pair changed dramatically when in close proximity. It was also concluded that the near field's contribution to the altered scatter properties was significant, and that it contributed to a symmetry break between the $m=\pm1$  scatter coefficients, even when the tubes were aligned along $x$. The near-field also altered the phase of the outgoing wave, demonstrating bizarre and unpredictable changes as the tubes were separated. Recently, a numerical study by \citet{felipe_2013} examined the interaction between thick flux tubes, finding the absorption of $m$ modes between the tubes to be effective in the multiple scattering regime. In contrast to \cite{hanasoge_2009}, they found that the multiple scattering regime can generate coherent phase change with separation distance, as well as a lack of symmetry breaking between the $\pm 1$ modes in the near field.

The aim of this work is to incorporate the sausage mode into the semi-analytical model of \citet{hanasoge_2009}, to investigate the interactions between the kink and sausage modes, and to settle the disparity with the numerical results of \citet{felipe_2013}.
This formalism is based on \citet{hanasoge_2009}, and uses the arguments presented by \citet{hindman_jain_2012} and \citet{andries_cally_2011} to implement the sausage mode. Extension of the method to collections of many close-packed disparate and arbitrarily situated tubes is straightforward, though at the expense of dealing with larger matrices, and will be explored in a future article. We will then be able to compare our calculations with seismic analyses of absorption and scattering by plage on the Sun \citep[e.g.,][]{braun_1995} to determine if the model captures the important characteristics of the interactions, and hopefully to draw conclusions about the nature of the constituent flux tubes.

Section~\ref{section:formalism} outlines the mathematical formalism for flux tube scattering, with Section~\ref{section:scatter} outlining the interaction with neighboring tubes. Section~\ref{section:results} outlines the results, with the relevance and comparison to \citet{hanasoge_2009} and \citet{felipe_2013} discussed in Section~\ref{section:discussion}.


\section{Mathematical Formalism}\label{section:formalism}

In this section we outline the interaction of flux tubes embedded in a stratified atmosphere of constant gravity.
 The field-free atmosphere is an adiabatically stratified ($\g =-2.775\,\times\,10^4\,\rm cm~\,s^{-2}\,\hat{\z} $)  truncated polytrope of index $m_p=1.5$.
 Following \cite{hanasoge_2009} we use an atmosphere with boundaries at $z_0=-392 $~km and $z=-98$~Mm. The pressure and the density of the atmosphere are
\begin{equation}
p(z)=p_0\left(-\frac{z}{z_0}\right)^{m_p+1}, \quad \text{and} \quad \rho (z)=\rho_0\left(-\frac{z}{z_0}\right)^{m_p}
\end{equation}
 respectively, where $p_0=1.21\,\times\, 10^5 \rm\,g\,cm^{-1}\,s^{-2}$ and $\rho_0=2.78\, \times\, 10^{-7}\rm\, g\, cm^{-3}$, we adopt a right-handed cylindrical coordinate system, where  $\x=(r,\theta,z)$. From this point on the index $m$ will denote the azimuthal order of the incident wave mode and $m'$ the scattered.
In this model a propagating $f$- or $p$-mode, with a vertical displacement eigenfunction $\Phi_m(\kappa^p_n;s)$, is given by;
\begin{equation}
 \Psi_{\rm inc}(\x,t)=\sum\limits^{n_p}_{n=0}\sum\limits^{\infty}_{m=-\infty}i^mJ_m(k^p_nr)\Phi_m(\kappa_n^p;s)e^{i(m\theta-\omega t)},
\end{equation}
where $J_m(z)$ is the Bessel function of the first kind of order $m$ and argument $z$.
 Defining a dimensionless depth $s=-z/z_0$, an $n$th order far-field eigenfunction is described by
\begin{equation}
\Phi_p(\kappa_n^p;s)=s^{-1/2-\mu}N_n\left[C_n^pM_{\kappa^p_n,\mu}\left(\frac{s\nu^2}{\kappa_n^p}\right)+M_{\kappa^p_n,-\mu}\left(\frac{s\nu^2}{\kappa_n^p}\right)\right],
\end{equation}
where $M_{\kappa,\mu}(z)$ is the Whittaker $M$ function of order $\kappa$ , $\mu$ and argument $z$, $N_n$ is the normalization constant for any  $p_n$ mode, and
\begin{equation}
\mu=\frac{m_p-1}{2},\;\;\nu^2=\frac{m_p\omega^2z_0}{g},\;\;k^p_n=\frac{\nu^2}{2\kappa^p_nz_0}
\end{equation}
are convenient dimensionless constants. The corresponding eigenvalues $\kappa_n$ are obtained through the relations outlined in Appendix A of \citet{hanasoge_etal_2008}. Here $n=0$ corresponds to the $f$-mode and $n\ge1$ to the $p_n$-modes.
The inclusion of gravity invokes the need to include not only a discrete set of far-field modes but also a continuous infinite set of near-field modes \citep{bogdan_cally_1995}. The near-field eigenfunction differs from the afore mentioned propagating eigenfunction through having complex roots:
\begin{equation}
\zeta_p(\kappa_n^J;s)=s^{-1/2-\mu}\left[C_n^JM_{-i\kappa^J_n,\mu}\left(i\frac{s\nu^2}{\kappa_n^J}\right)+M_{-i\kappa^J_n,-\mu}\left(i\frac{s\nu^2}{\kappa_n^J}\right)\right]
\end{equation}
In truncating the polytrope, much like \citet{Barnes_Cally_2000}, at $98$~Mm we reduce this continuous spectrum of jacket modes to an unphysical discrete spectrum. This truncation is required in order to utilize the interaction theory of \citet{kagemoto_yue_1986}. Selecting a large enough discrete set of jacket modes, we mimic the true jacket spectrum and develop some understanding of the nature of near-field interactions. Selecting a larger set (500 more) of jacket modes than the one we have used, alters the results by less than a percent.


\subsection{Thin Flux Tubes}
The determination of the scattering matrix for a single tube is essential in trying to understand the interaction between nearby flux tubes. Consider a flux tube that is embedded in a field free atmosphere that responds to an incident $f-$mode. At depths below $s=1$ the filament's radius is small when compared to the incident wavelength. The relatively small radius allows for the utilization of the thin flux tube approximation, as outlined by \cite{bogdan_etal_1996}. In this approximation, the longitudinal magnetic field $b$ and the tube radius $R$ are given by
\begin{equation}
b(z)\approx \sqrt{\frac{8\pi p(z)}{1+\beta}},\quad \textrm{and} \quad \pi R^2(z)\approx\frac{\Phi_f}{b(z)}\,,
\end{equation}
where $\Phi_f =3.88\times10^{17}~\rm Mx$ is adopted as the total magnetic flux per tube throughout this article.
Rapid expansion of the flux tubes above this height requires both the introduction of complex higher order fluting modes, as well as a variable plasma-$\beta$ across the tubes' cross-sections. In the thin tube approximation their slender nature forces $\beta$ to be constant throughout the tube, as it is in thermal and radiative equilibrium with the field free atmosphere. This is distinct from the thick tubes of \cite{felipe_2013}, where $\beta$ varies both along and across the tubes.

\begin{figure*}
\includegraphics[width=\linewidth]{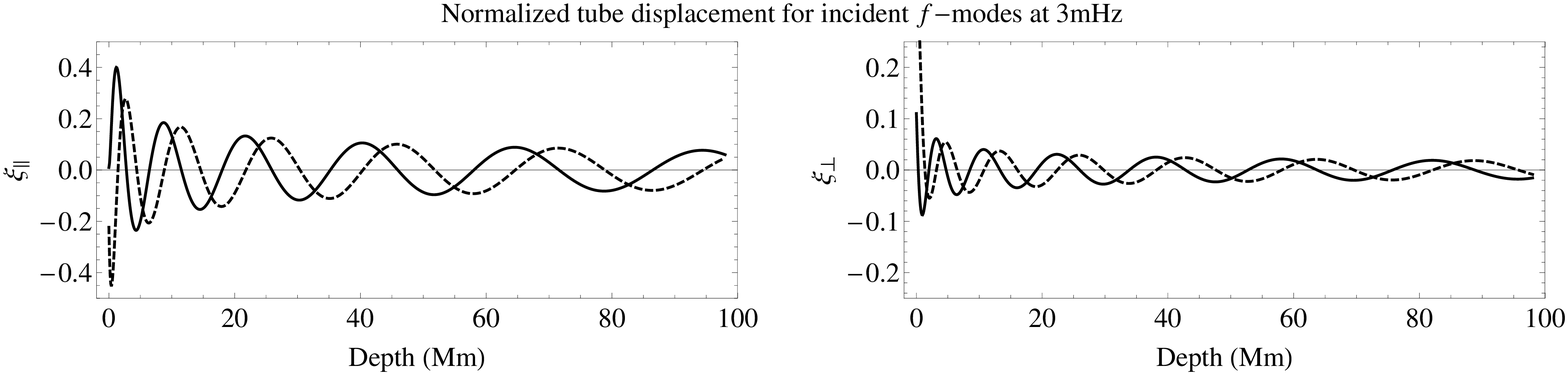}
\caption{ The real (solid) and imaginary (dashed) components of the normalized vertical and radial displacements ($\xi/k_p$), of a flux tube ($\beta=1$) that is impinged upon by a 3~mHz $f$-mode of order $m=0$ (left) and $m=1$ (right). The motions are vertical and radial for the sausage and kink modes respectively. The slow waves traveling down the flux tube propagate into the solar interior, transporting energy absorbed from the incident wave.}
\label{fig:displacement}
\end{figure*}

Thin magnetic flux tubes are only capable of two oscillating modes, the $m=0$ and $m=\pm1$, where $e^{im\theta}$ dependence is assumed. Incident waves, $\Psi_{\rm inc}$, of these $m$ mode orders generate horizontal and vertical displacement within the tube (Figure~\ref{fig:displacement}), resulting in the tube oscillating in a sausage ($m=0$) or the kink ($m=\pm1$) like motion according to
\begin{equation}
\left[ \omega^2(2m_p+\beta(m_p+1)) + \frac{2gs}{z_0}\frac{\partial^2}{\partial s^2} + \frac{g(m_p+1)}{z_0}\frac{\partial}{\partial s} \right]\xi_{\|} = -\omega^2(m_p+1)(\beta+1)\frac{\partial \Psi_{\rm inc}}{\partial s}
\end{equation}
and
\\
\begin{equation}
\left[ \omega^2 z_0 +\frac{2gs}{(1+2\beta)(m_p+1)} \frac{\partial^2}{\partial s^2}+\frac{g}{1+2\beta} \frac{\partial}{\partial s} \right]\xi_{\perp}=
\frac{2(1+\beta)}{1+2\beta}\omega^2z_0\frac{\partial \Psi_{\rm inc}}{\partial x}
\end{equation}
respectively \citep{bogdan_etal_1996}, where $\omega$ is the angular frequency of the incident wave.
In terms of the energy budget for the system, not all of the energy of an incident wave is transferred to vertically propagating slow waves. Some of the energy is scattered by the flux tube into the external medium. This scattered wave field is determined through the matching of the internal motion and the pressure to those of the external medium. Thus, upon responding to an incident wave, tube $i$'s scattered wave is:
\begin{equation}\label{eq:scatter}
\phi^S_i(r_i,\theta_i,s)=-\sum\limits_{m=-1}^{1}\left[ \sum\limits_{n=0}^{n_p} S^i_{mn}\Phi_n(\kappa_n^p;s)H^{(1)}_m(k_n^pr_i)e^{im\theta_i} + \sum\limits_{n=n_p}^{N} S^i_{mn}\zeta_n(\kappa_n^j;s)K_m(k_n^jr_i)e^{im\theta_i}  \right ],
\end{equation}
where $H^{(1)}_m(z)$ and $K_m(z)$ are the Hankel and K-Bessel functions of order $m$ and argument $z$.
The scattering matrix $S^i_{mn}$contains all the scattering coefficients of the propagating and evanescent modes. These coefficients are found through the mismatch between the internal and external wave fields. 

Satisfactory boundary conditions must be maintained in the calculation of the scattering coefficients. This requirement is that the pressure and horizontal displacement must match at the tube boundary. To leading order of tube radius $R$, as $R\rightarrow 0$, the pressure is of higher order (for all three wave fields: internal, scattered and incident) compared to that of the displacement for $|m|=1$. The agreement at the tube boundary is then achieved, to leading order, through the matching of the horizontal displacement $\xi_\perp$ alone:
\begin{equation}
 S^{i}_{\pm1n} \xi_\perp^{\text{scat}} =\xi_\perp^{\text{int}}-\xi_\perp^{\text{inc}},
\end{equation}
 where scat, int and inc are the scattered, internal and incident wave fields respectively. Whilst the matching of the kink modes is achieved relatively easily, the matching of the sausage modes has presented problems in the past. Unlike the kink mode matching, the pressure terms are not small, resulting in both the pressure and displacement continuity needing to be maintained for all three wave field components.
The complexity of matching all three terms in the thin tube approximation has restricted previous studies. However, the requirements and achievement of matching have recently been outlined by \cite{hindman_jain_2012} and \cite{andries_cally_2011} concurrently, through the comparison of small argument expansions (as $R(z)\ll \lambda$) of the Bessel functions. The internal and incident pressures can be matched to leading order $R$, since the scatter pressure terms are proportional to $\ln R$. To leading order, matching the internal and incident wave fields maintain the pressure continuity. In turn, to leading order, the mismatch is then calculated through the matching of the internal  normal displacement ($O(R)$) to that of the incident ($O(R)$) and scattered ($O(R^{-1})$) components. As long as the the scatter terms are proportional to  $R^2$, the displacement continuity is maintained. Using these arguments, the scatter coefficients for the sausage mode are then calculated through:
\begin{equation}
D_n S^i_{0n} \phi_n=\frac{R}{z_0^2}\left[N_{\text{inc},n}(\omega;z)-N_{\text{int} ,n}(\omega;z)\right],
\end{equation}
where $D_n$ is $-2i/\pi$ for $n\le n_p$, and $-1$ for $n>n_p$ and $N=\tilde{\mathbf{n}}.\grad  \Phi$ is the normal displacement \citep{hindman_jain_2012}.

The absorption and phase shift of waves encountering magnetic regions on the Sun's surface can be quantified through Hankel analysis \citep{braun_1995}, whereby ingoing and outgoing waves in an annular pupil surrounding an active region or plage are compared. Plage in particular is made up of very many separate flux tubes packed randomly but closely in an extended region. For us to make best use of Hankel data therefore requires the development of multiple scattering theory including near-field effects. Although we only treat two tubes here, extension to many tubes will follow. With these tools, we hope to be able to probe the nature of plage and its constituent flux tubes using observed absorption and phase shifts.

We define the absorption of an $(m,n)$ incident wave in the usual manner,
\begin{equation}\label{eq:abs}
\alpha_{mn}=\frac{|A_{\textrm{in}}|^2-|A_{\textrm{out}}|^2}{|A_{\textrm{in}}|^2},
\end{equation}
where $A_\textrm{in}$ and $A_\textrm{out}$ are the complex amplitudes of the incident and scattered waves respectively. The change in phase of the scattered wave is
\begin{equation}\label{eq:phase}
\Delta\phi_{mn}=\arg\left\{\frac{A_{\textrm{out}}}{A_{\textrm{in}}}\right\}.
\end{equation}
The `absorption' so defined includes both true absorption by the tubes, and scatter into other $(m',n')$ outgoing waves. A single (circular) tube does not scatter in $m$, though it can in $n$. Multiple tubes though will scatter in $m$ as well. The energy scattered into these other modes is best quantified in terms of energy fractions. Generalizing the analysis of \citet[see their Eq.~(27)]{hindman_jain_2012}, the fraction of the incident wave's energy that is scattered to outgoing $m'$ and $n'$ is 
\begin{equation}\label{eq:energyfraction}
\epsilon_{mn\rightarrow m'n'}=|\delta_{nn'}\delta_{mm'}+2S_{m'n'}|^2,
\end{equation} 
where $S_{mn}$ are the scattering coefficients ($n\le n_p$) appearing in Equation (\ref{eq:scatter}). 


\section{Interaction of flux tubes}\label{section:scatter}
\begin{figure}
\centering
\includegraphics[scale=0.7]{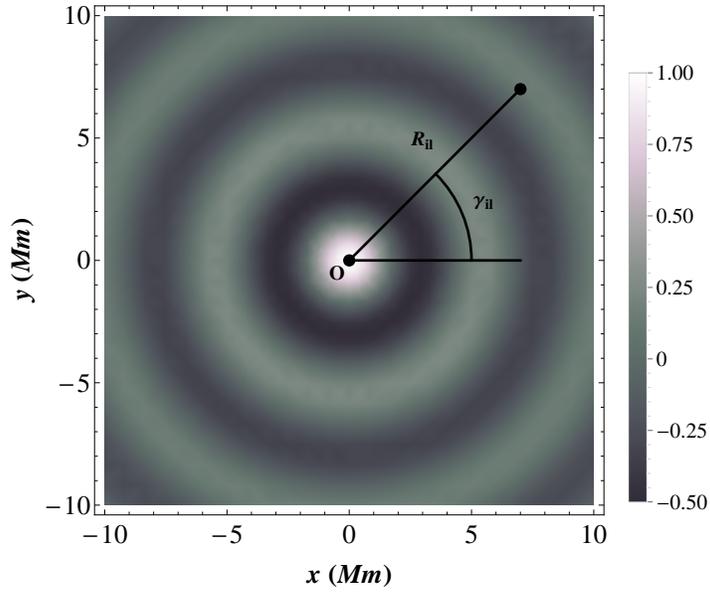}
\caption{The orientation of a pair of flux tubes with variables graphically specified. The two tubes are separated by distance $R_{il}$, with the line at an angle of $\gamma$ from the $+x$ axis. An incident sausage mode ($J_0(k_pr)$ is centered upon one of the tubes, whilst the off-center tube sees less of the incident wave the further away it is.}
\label{fig:orient}
\end{figure}
Having determined the scatter coefficients for a single tube $i$, we now build upon this matrix to include not only the incident wave but also the contribution of nearby tubes. Using the methodology of \cite{kagemoto_yue_1986}, the scattered wave (Equation~\ref{eq:scatter}) from an isolated tube $i$ is expressed, in matrix notation, as
\begin{equation}\label{eq:sing}
\phi^S_i=\sum\limits_nA_{in}^T\Psi_{in}^S,
\end{equation}
where the summation is over all propagating and evanescent modes. In the summation the jacket modes follow on from $n_p$. Here $A_{in}$ is a 3-element vector containing the scattering coefficients for the sausage and kink modes, whilst $\Psi_{in}$ is a vector containing the scattered wave field from tube $i$.
Specifically:
\begin{equation}
A_{in}=-\begin{pmatrix}S_{-1n}^i & S_{0n}^i & S^i_{1n}\end{pmatrix}^T
\end{equation}
\begin{equation}
(\Psi^S_{in})_{cd}=\begin{matrix}H^{(1)}_{c-2}(k^p_nr_i)\Phi_n(\kappa_n^p;s_d) & & (n\le n_p) \end{matrix}
\end{equation}
\begin{equation}
(\Psi^S_{in})_{cd}=\begin{matrix}K_{c-2}(k^p_nr_i)\zeta_n(\kappa_n^p;s_d) & & (n> n_p), \end{matrix}
\end{equation}
where $c$ ranges over $[1,3]$, $d$ over $[1,250]$, and $s_d$ is the $d^{th}$ point along the $s$ grid.
Equation~(\ref{eq:sing}) describes only the zeroth order  incident wave ($\phi_0$) contribution. The scattered wave contribution from nearby tubes is then described as an incident wave upon tube $i$, and this is achieved through the \emph{Transformation Matrix}. 
The transformation matrix $\T_{il}$ relates the scattered wave field from tube $i$ to an incident wave upon tube $l$, that has a  radial separation of $R_{il}$ and angular separation from the $x$ axis of $\gamma_{il}$ (Figure~\ref{fig:orient}). The elements of $\T_{il}$ are derived from  Graf's addition formula \citep{abram_1964}:
\begin{equation}
H^{(1)}_m(k_n^pr_i)e^{im(\theta_i-\gamma_{il})}=
\sum\limits^{\infty}_{d=-\infty}H_{m+d}^{(1)}(k_n^pR_{il})J_d(k_n^pr_l)e^{id(\pi-\theta_l+\gamma_{il})},
\end{equation}
\begin{equation}\label{eq:fix}
K^{(1)}_m(k_n^jr_i)e^{im(\theta_i-\gamma_{il})}=
\sum\limits^{\infty}_{d=-\infty}K_{m+d}^{(1)}(k_n^jR_{il})I_d(k_j^pr_l)e^{id(\pi-\theta_l+\gamma_{il})}.
\end{equation}
The individual elements then populate $\T_{il}$ through
\begin{equation}
(\T_{il})_{pq}=e^{i(q-p)\alpha_{ij}}H^{(1)}_{q-p}(k_pR_{il}) \quad (n\le np),
\end{equation}
and
\begin{equation}\label{eq:jacT}
(\T_{il})_{pq}=(-1)^{p}e^{i(q-p)(-\pi/2+\gamma_{ij})}K_{q-p}(k_jR_{il})  \quad (n > np)
\end{equation}
where in this case  $p=3n+m'+2$ and $q=3n+m+2$. Close inspection of the exponential in Equation~(\ref{eq:jacT}) reveals that it is different from that of \citet{kagemoto_yue_1986} (as an $e^{-\pi(q-p)/2}$ term is introduced), as well as \citet{hanasoge_2009}. This additional factor has been introduced because of the apparent break in symmetry between the $m=\pm$1 in the near-field, and effectively correspond to representing the solution in terms of a Hankel function, instead of a Bessel $K$ \citep[see][Equation 9.6.4]{abram_1964}. \cite{hanasoge_2009} concluded that jacket modes create a break in symmetry in the scattering coefficient between the $m=\pm$1 modes, when the tubes are aligned along $x$. However, the only difference between the m=$\pm1$ modes is the definition of the coordinate system. If the coordinate system were reflected around the $x$ axis, the scattering coefficient should not change. Any tube located at the origin of the coordinate system should not be able to identify a difference in scatter between $m=\pm1$ from the off-center tube when it is on the $x$-axis. The apparent break in symmetry is significant in the near-field, but this is due to the chosen solution describing the jacket and incident wave fields described in \cite{kagemoto_yue_1986}. We will further address the mathematics of this issue later on, once we have outlined the remaining matrices needed in the calculation of the scatter.

By using $\T_{il}$ to define the scatter as an incident wave, Equation~(\ref{eq:scatter}) can then be expressed for any neighboring tube $l$ as
\begin{equation}\label{eq:20}
\phi_l^I=\sum\limits_n\left(\phi_{0}|_{ln}+\sum\limits_{i=1,i\ne l}^{N}A_{in}^T\mathbf{T}_{il}^n\Psi_{ln}^I\right),
\end{equation}
where the interior sum describes the contribution of all other tubes.
 To better understand the contribution of all other tubes, the scatter and incident waves must be related through the characteristics of the isolated tube \citep{kagemoto_yue_1986}. As such there exists diffraction transfer matrices, called here the $\B$ matrix, that relates the incident and scattered wave field for tube $l$:
\begin{equation}\label{eq:21}
\A_l=\B_l\a_l,
\end{equation}
where $\B$ is populated through
\begin{equation}
(\B)_{pq}=S_{mn}|_{n'}.
\end{equation}
In this case $p=3n'+m+2$, $q=3n+m+2$, the vector $\a_l$ contains the amplitude of the incident $n$ wave upon tube $l$ and $\A_l$ contains the scatter terms into all $n$. The incident wave amplitudes ($\a_l$) are again derived from Graf's addition formula,
\begin{equation}\label{eq:incid}
J_m(k_n^pr_i)e^{im(\theta_i-\gamma_{il})}=
\sum\limits^{\infty}_{d=-\infty}J_{m+d}(k_n^pR_{il})J_d(k_n^pr_l)e^{id(\pi-\theta_l+\gamma_{il})},
\end{equation}
noting that an $m$ incident wave will be seen as a combination of all other $m$ by any tube that is not located at the origin. It is important to note that the axisymmetric flux tubes create a sparse $\B$, with no information pertaining to scatter into non incident $m$ modes. Hence, all $m$-$m'$ scatter is due to off-center tubes experiencing the incident wave as a combination of $m$ modes.

As with the $\T$ matrix, the $\a$ vector is constructed through Graf's addition formula, and herein lies the symmetry problem. The symmetry break arises due to the nature of the $J_m$, $H^{(1)}_m$ and $K_m$ functions, of positive and negative $m$ order. For odd $m$ the Bessel $J$ (as well as $H^{(1)}_m$) function behaves like $J_{-m}=-J_m$, and as a result through solving the linear algebra for $\A_l$ no difference arises between $\pm m$ . As the far-field components are described by $J_m$ and $H^{(1)}_m$, the matrix procedure of \cite{kagemoto_yue_1986} then sees no difference in the scattered far-field $\pm m$ waves (when the tubes are aligned with $x$). However, the Bessel $K$ function is invariant for $\pm m$, yet when multiplied by the variant  $J_m$ (which is present in $\a_l$), terms do not cancel, and thus a difference in the $\pm m$ scattering coefficient arises.
By examining the definition of the Bessel $K$ function \citep[see][Equation 9.6.4]{abram_1964} the exponential factor returns the symmetry of the $\pm 1$ modes when the tubes are aligned along the $x$ axis, whilst remaining an appropriate solution. 

Expanding upon Equation~(\ref{eq:20}), by utilizing Equation~(\ref{eq:21}), completes the picture of the scattered wave field from a flux tube pair:
\begin{equation}\label{eq:scatmatr}
\A_l=\B_l\left(\a_{l}+\sum\limits_{i=1,i\ne l}^{N}\mathbf{T}_{il}^T\A_l\right),
\end{equation}
\begin{equation}\label{eq:Aone}
\A_1=\B_1(\a_1+\T^T_{21}\B_2[\a_2+\T^T_{12}\A_1]),
\end{equation}
\begin{equation}\label{eq:Atwo}
\A_2=\B_2(\a_2+\T^T_{12}\B_1[\a_1+\T^T_{12}\A_2]).
\end{equation}

\section{Results}\label{section:results}
\begin{figure*}
\centerline{\includegraphics[height=17cm]{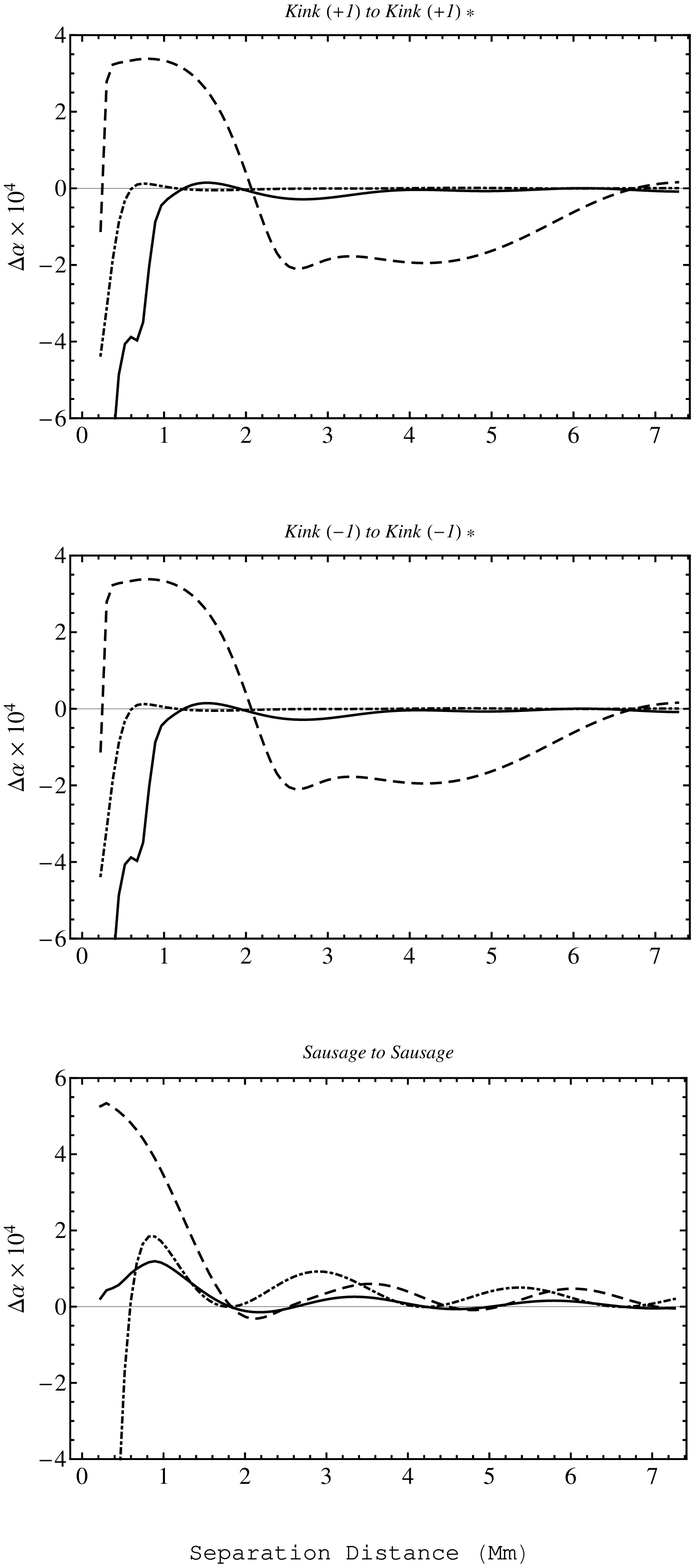}}
\caption{The change in absorption coefficient ($\Delta\alpha=\alpha-\alpha^{\rm isolated}$) (of the central tube only) of an incident $m$ wave due to the presence of a nearby tube for varying separation distances and $\beta$. The $3$~mHz ($\lambda=4.9$ Mm) $f$-mode interacts with an identical pair of flux tubes, aligned along the $x$-axis ($\beta=0.1$: Dashed, $\beta=1$: Solid, $\beta=10$: Dot Dashed).  Frames specified by * have the $\beta=0.1$ absorption scaled down by factor 10 for clarity in near-field contribution. The absorption coefficients are calculated from Equation~(\ref{eq:abs}). In all panels the central tube's absorption is dramatically altered (compared with the single tube absorption listed in Table~\ref{table:isoabs}) when interacting with the nearby tube through the near-field.}
\label{fig:mmdashkink}
\end{figure*}

Let us consider the case of two identical tubes (same $\beta$ and $\Phi_f$), aligned along the $x$ axis and interacting through the sausage and kink modes. For this study we concern ourselves only with the $f$-$f$ scattering coefficients, as $p_n$ scatter decreases rapidly with increasing $n$ \citep{bogdan_etal_1996}. Upon scattering an incident wave, each tube will also experience the scatter of the other tubes, and in turn produce further scatter. The resultant changes in absorption for the central tube are seen in Figure~\ref{fig:mmdashkink}, for a mode ($m$) scattering into an outgoing mode ($m$).  The greatest changes in absorption (on the order of $10^{-4}$), for both modes, occur when the tubes are interacting through the near field, typically on a length scale of half the horizontal wavelength ($\pi / k_p$). However, the kink modes demonstrate a sensitivity to the tube's plasma-$\beta$,  with smaller $\beta$ tubes readily absorbing more (a magnitude greater) than higher $\beta$ tubes. Comparison of these changes with the isolated case (Table~\ref{table:isoabs}), demonstrates that the presence of the second tube alters the isolated absorption coefficient significantly.
\begin{deluxetable}{cccc}
\tablecaption{The absorption coefficient for an isolated tube at the origin\label{table:isoabs}}
\tablecolumns{4}
\tablewidth{0pt}
\tablehead{$|m|$&$\beta=0.1$&$\beta=1$&$\beta=10$}
\startdata
0 & $2.16 \times 10^{-4}$ & $2.45 \times 10^{-3}$ & $6.34 \times 10^{-3}$ \\ 
$1$ & $3.55 \times 10^{-2}$ & $1.15 \times 10^{-2}$ & $1.18 \times 10^{-3}$ \\ 
\enddata
\end{deluxetable}

\begin{figure*}
\centerline{\includegraphics[height=17cm]{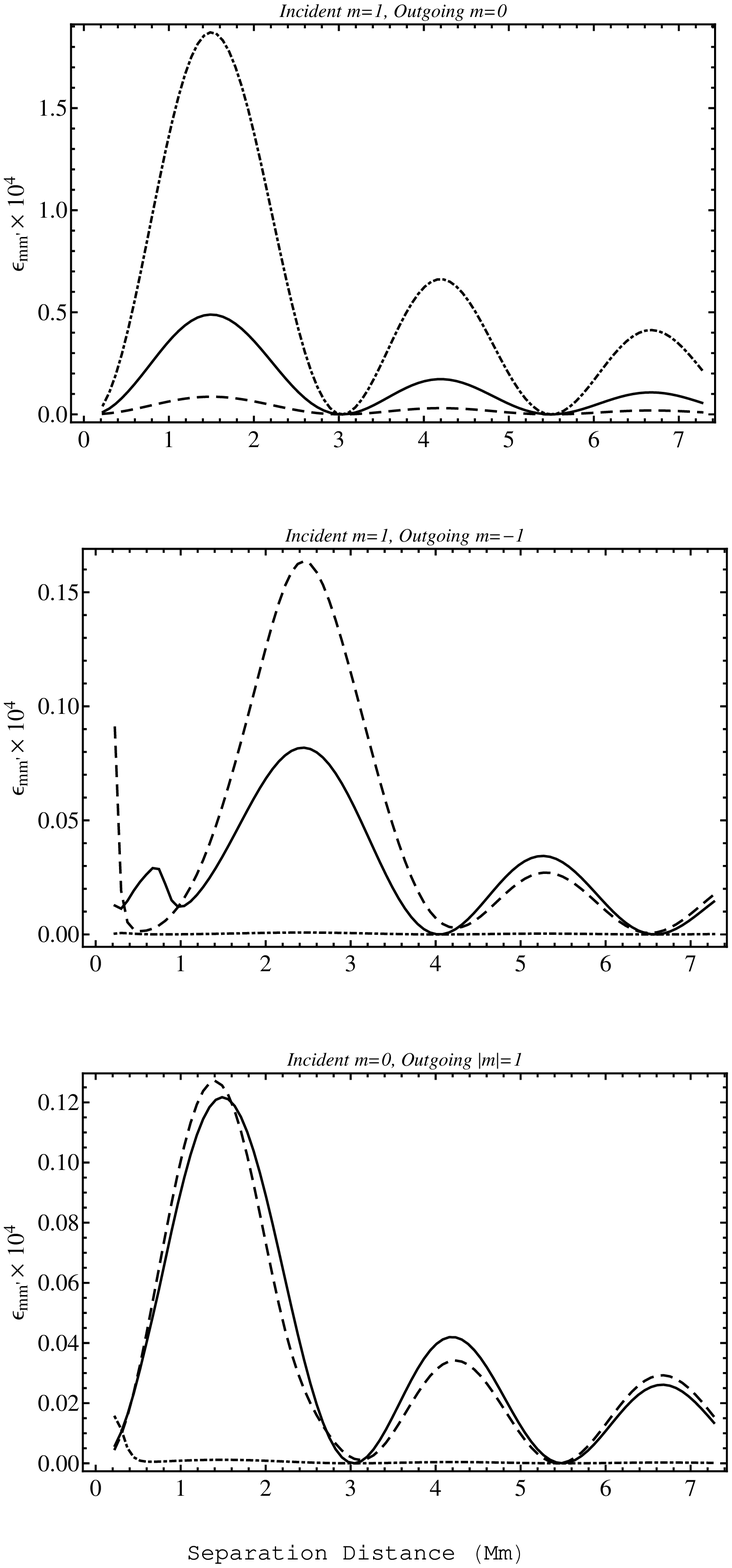}}
\caption{The fraction of energy ($\epsilon_{mm'}$) that is scattered into non-incident $m'$ modes by a pair of identical tubes for varying separation distances and $\beta$. The $3$mHz ($\lambda=4.9$Mm) $f$-mode interacts with an identical pair of flux tubes, aligned along the $x$-axis ($\beta=0.1$: Dashed, $\beta=1$: Solid, $\beta=10$: Dot Dashed). All $\beta=0.1$ energy fractions are scaled down by factor 10 for clarity of the near-field contribution.  Scattering into kink modes is strongest when the tubes have a small plasma-$\beta$. The energy fraction is calculated from Equation~(\ref{eq:energyfraction}).} 
\label{fig:mmdashsaus}
\end{figure*}

Scattering into nonincident ($m'\ne m$) modes  is the subject of Figure~\ref{fig:mmdashsaus}. The fraction of energy carried away by these $m'$ modes is greatest when the tubes are close enough to interact through their near-fields. However, as they are separated the energy fraction decreases.  The energy fraction also diminishes as the tube separation approaches zero, due to the system returning to an axisymmetric state.

\begin{figure*}
\includegraphics[width=\textwidth]{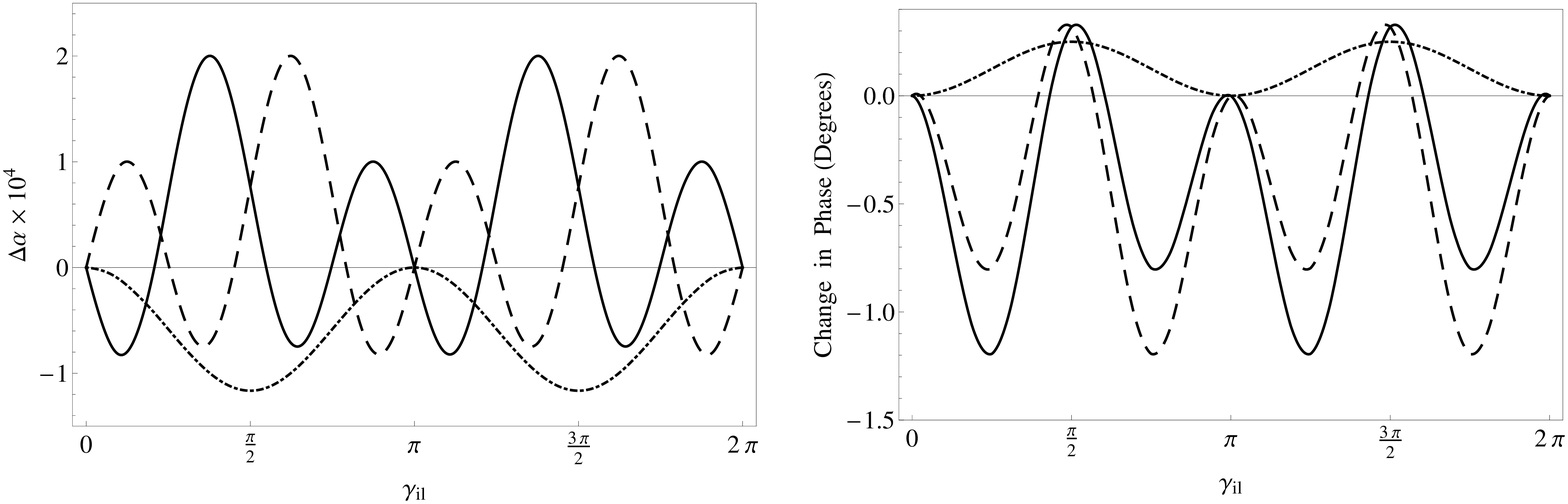}
\caption{The change in absorption and phase of the central tube, as the second tube is rotated around the origin at a distance of 500~km. The incident $m$ modes (solid: +1, dashed: -1, dot-dashed: 0) scatter off the second tube producing a change in the central tube. Symmetry between the $\pm 1$ modes is restored when the second tube is aligned on the $x$-axis ($\gamma_{il}=0,\pi$). Whilst the $m=0$ mode is axisymmetric a change occurs when the second tube is rotated about the origin, due to the kink mode scatter being absorbed as sausage mode components (see Equation~\ref{eq:incid}).}
\label{fig:dgamma}
\end{figure*}

In this model, we have applied a different solution for the jacket modes (Equation~\ref{eq:fix}) in an attempt to restore symmetry between $\pm m$ modes.
As a result the apparent symmetry breaking between $m=\pm1$ \citep{hanasoge_2009}, when the tubes are aligned along $x$, is absent. Scattering into their respective modes is identical for both the $+1$  and $-1$ modes (see two top panels of Figure~\ref{fig:mmdashkink}). However, as the second tube is rotated about the origin the symmetry is broken and a difference in absorption arises between the $-1$ and $+1$ modes (Figure~\ref{fig:dgamma}).  In regards to the sausage mode, this mode naturally has no preference between which kink mode to scatter into (see bottom panel of Figure~\ref{fig:mmdashsaus}).  

\begin{figure*}
\centering
\includegraphics[width=17cm,height=10cm]{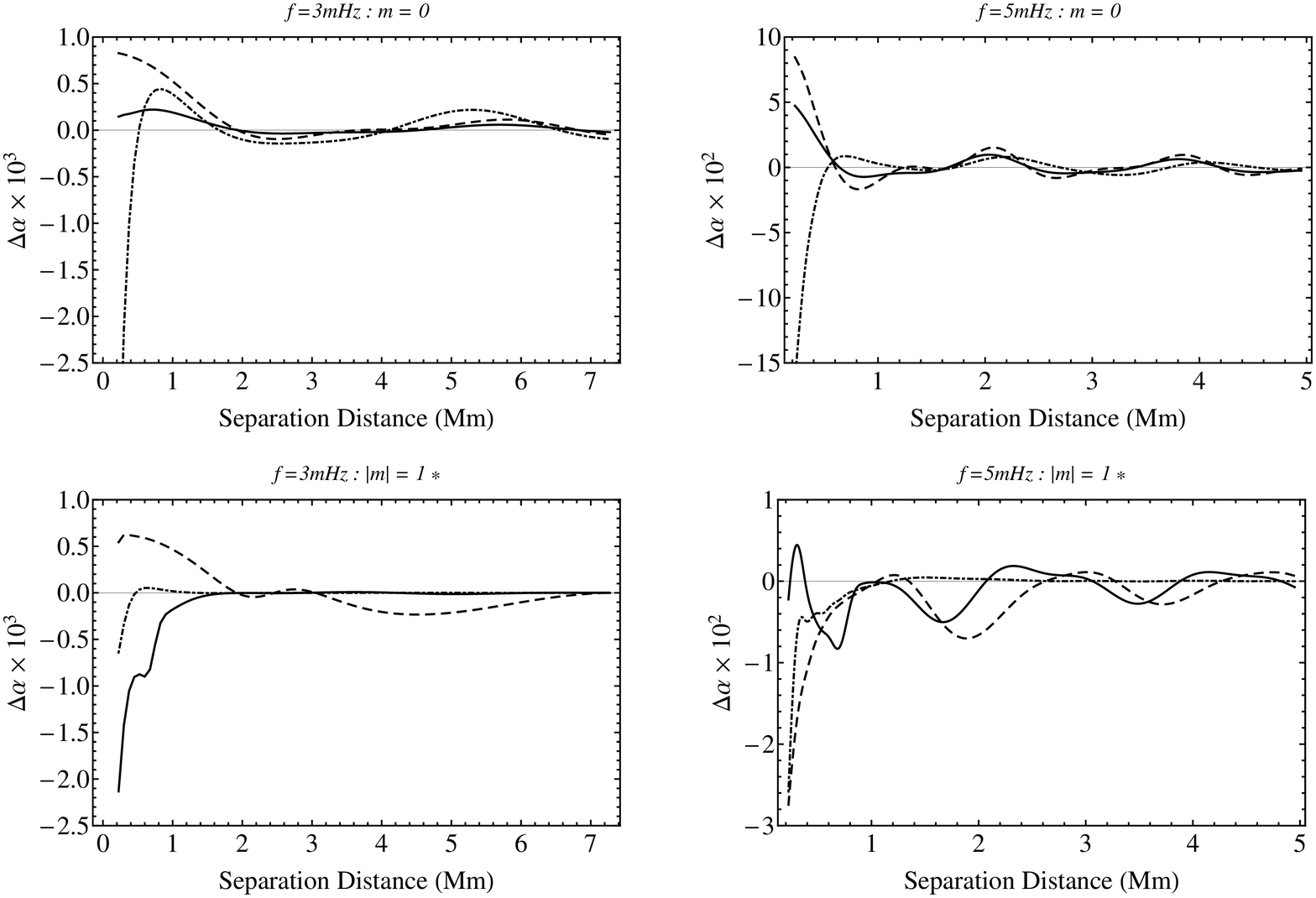}
\caption{The change in absorption ($\alpha-\alpha^{\rm isolated pair}$) of an identical pair aligned along the $x$-axis. Absorption is generally enhanced within the near field, slowly returning to isolated values as the tubes are separated. Left Panels: Comparison with Table~\ref{table:isoabs} reveals the absorption of the pair is enhanced compared to that of the central tube. The pair behave collectively. Line types are as in Figures \ref{fig:mmdashkink} and \ref{fig:mmdashsaus}.}
\label{fig:mmscatter}
\end{figure*}

\begin{figure*}
\centering
\includegraphics[width=17cm,height=10cm]{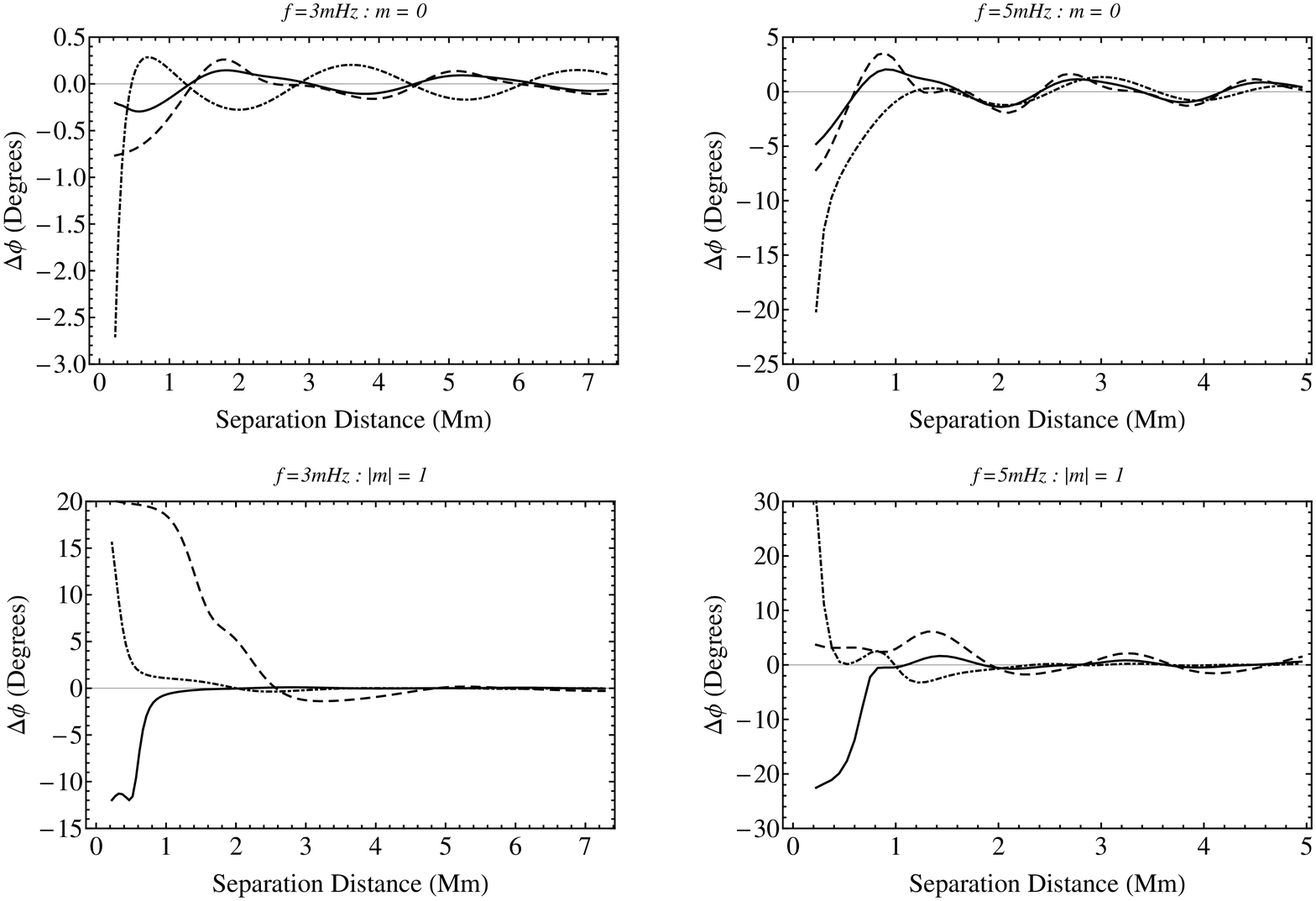}
\caption{The change in phase of the scattered wave field from the pair, compared to that of two isolated tubes. Large phase changes are seen in the near-field, rapidly returning to isolated values as the tubes are separated beyond a wavelength of the incident mode. Kink mode scattering (bottom panels) produces the largest change in phase with high frequency also contributing the increased phase changes. Smooth changes in phase occur as the tubes are separated, contrasting the results of \cite{hanasoge_2009}. Line types are as in Figure~\ref{fig:mmdashkink} and \ref{fig:mmdashsaus}.}
\label{fig:dphase}
\end{figure*}

Expanding our view to the complete scattered wave field, rather than that of the central tube,   reveals the collective nature of a pair of flux tubes.  Akin to the central tubes absorption (and \cite{hanasoge_2009}, despite the change to the jacket solution), the absorption of the flux tube pair  varies greatly when allowed to interact through their respective near fields (Figure~\ref{fig:mmscatter}). The changes in absorption are again significant when compared to the non-interacting tube's absorption coefficient (Table~\ref{table:isoabs}), returning to an isolated absorption coefficient when sufficiently far apart. At higher frequencies, incident waves are  more readily absorbed by the tubes. However, due to a shorter wavelength the pair need to be closer (in comparison to lower frequencies), to interact through their respective near fields. The affects of the  proximity thus manifests itself in an altered absorption coefficient, and hence the tubes no longer absorb and scatter as two unique isolated tubes but rather as a collective scatterer.
The impact of this collective behavior is also evident in the phase of the outgoing scattered wave field (Figure~\ref{fig:dphase}). The sensitivity of the phase with tube separation \citep{hanasoge_2009} is not present here, with coherent changes in phase as the tubes are separated. The greatest changes in phase generally occur in the kink modes with the higher frequency incident waves also aiding in an  increased change in phase. The near field interactions are the cause of the greatest phase changes, with the impact on travel times increasing with proximity.

\section{Discussion and Conclusions}\label{section:discussion}
Whilst the results presented here are informative about $m$ mode interactions, we must first discuss limitations of the model before considering their significance.  
In truncating the polytope we reduce the number of jacket modes from a continuous spectrum to a discrete one. The nature of the finite-depth interaction theory of \cite{kagemoto_yue_1986}, requires the use of this discrete spectrum, but how much does the scatter differ when the polytrope is allowed to extend to infinite depth? In numerical simulations the jacket modes appear naturally, but these simulations are restricted in spatial resolution. By treating the jacket eigenfunctions analytically, we are not as easily restricted by spatial resolution, and can consider their impact. In taking a large enough subset of the jacket modes the eigenfunctions of the tube are mimicked. However,  attributing results from a partial basis to a complete picture must be done cautiously. 
The second limitation is the use of the thin tube approximation. Whilst completely valid below $z_0$, how much are we underestimating the degree of scatter from higher order fluting modes present above $z_0$? \citet{felipe_2013} performed numerical studies of thick interacting tubes (capable of $m\ge2$ modes), showing that whilst the sausage and kink modes are the dominant modes for scatter, the contribution of higher modes is not negligible. Recognizing this, we note that this simplified model is not perfect in modeling realistic interactions between tubes that have a comparable thickness to the incident wavelength. Nevertheless, it is a helpful tool for parameter searches to characterize the multiple scattering regime.

Having stated these limitations let's consider our adjusted near field solution.
As discussed previously, the difference between $m=\pm1$ modes is the definition of the coordinate system. The near field solution of \citet{kagemoto_yue_1986} presents a symmetry problem for tubes interacting through $\pm m$ modes. Whilst the Bessel $K$ solution for the scattered jacket modes is accurate, the invariant nature of $K_m$ (in contrast to $H^{(1)}_m\text{ and } J_m$) between $\pm m$ leads to symmetry breaking in the calculation of Equation~(\ref{eq:Aone}). 
Our Bessel $K$ solution (Equation~\ref{eq:fix}), which is a valid solution for the jacket modes, has restored the scatter symmetry when the tubes are aligned along $x$. This solution maintains the sensibility that $m=\pm1$ modes differ only through a coordinate system definition, and not through their respective scatter. Consequentially, without this new solution the interaction theory would stipulate that the $m=0$ mode, being axisymmetric, would show preferential scatter into $\pm m$ modes. We also note that the incoherent phase changes as the tubes are separated, found by \citet{hanasoge_2009}, are absent in these results. This is again due to the previous near-field solution. Our results show coherent changes in phase  as the tubes are separated, which is consistent with the changes in phase of \citet{felipe_2013}.

Interestingly, by moving the off-center tube to a position where $\gamma_{il}\ne0$, the symmetry is no longer maintained (Figure~\ref{fig:dgamma}). The  $e^{im\theta}$ dependence of $m\ne0$ waves creates a symmetry break between modes (when $\gamma_{il}\ne0$), owing to unequal amounts of $\pm m$ waves at the second tube. Consequentially, a difference in the scatter coefficient of the $m=\pm1$ modes arises at the second tube, and in turn a difference at the centered tube also occurs. The sensitive nature of the scatter, to the tube position, was also observed by \citet{felipe_2013} in numerical studies for incident plane waves. This sensitivity outlines another parameter to consider in the determination of a ensemble's constitution. Changes in the sausage mode occur when the second tube is rotated about the origin, due to the second tube scattering kink modes, and in turn the central tube absorbs $m=0$ modes (as per Equation~\ref{eq:incid}). The absorption of all modes is identical if the origin is relocated to the center of the second tube (i.e. $\gamma_{il}=\pi$), and this serves as a sanity check for the coordinate system definitions.

The inclusion of the sausage mode in this model highlights it's importance in the multiple scattering regime. The change in absorption for the sausage mode is of the same magnitude as the kink modes for the 3 mHz case (Figure~\ref{fig:mmdashkink}), owing to the fact that the off center tube sees any incident $m$ wave as a combination of $m$ modes, and in turn scatters significantly into both sausage and kink. However, in the higher frequency (5 mHz) case the absorption of the sausage mode is a magnitude larger than that of the kink modes when the tubes are in close proximity (see right panels of Figure~\ref{fig:mmscatter}). These results are consistent with \citet{felipe_2013}, which illustrated that high frequency cases will produce larger absorption changes for the $m=0$ modes than for the $|m|\ge1$ modes. With respect to the phase shift of the individual $m$ modes (Figure~\ref{fig:dphase}) the close proximity of tubes generally enhances the negative phase shift of the sausage mode \citep{hindman_jain_2012}, which results from shorter travel distances due to reflection from tube boundaries. Akin to the absorption, the phase shift of the sausage modes is enhanced by the higher frequency incident waves becoming comparable to the phase shift generated by kink modes. Figure~\ref{fig:dphase} illustrates that close proximity and higher frequency incident waves may enable the phase shift produced by $m=0$ waves to surpass that of the kink modes in some cases (such as thicker tubes), as anticipated by \citet{felipe_2013}.  

The desire to determine the properties of magnetic ensembles from the scattered wave field results from the inability to directly discern the internal constitution of magnetic features. Techniques to determine the amplitude and phase of scattered waves from small magnetic elements from observations exist (see \cite{duvall_etal_2006}), but which parameters govern the mechanisms that affect the scattered wave field? The multiple scattering regime is well known to alter the scattered wave field \citep{bogdan_zweibel_1987, keppens_etal_1994}, and cannot be approximated by single scattering. Numerical studies have begun to characterize this regime \citep{felipe_2013}, and we have presented here a more complete semi-analytical model to aid in the determination of parameters that govern the scattered wave field, demonstrating that it is sensitive to both the proximity and relative position of nearby tubes. The inclusion of the sausage mode has also illustrated the significance of the scattering between $m$ modes, especially within the near-field. In more detailed cases the model presented here can be used to address the scattering between thin tubes, like those present within plage. In a future study, we will apply the formalism developed here to address the interactions of multiple non-identical tubes in random ensembles.

The authors thank Shravan Hanasoge for his insightful conversations and discussions on the formalism.

\bibliographystyle{apj}        
\bibliography{References}
\end{document}